\documentclass[11pt]{elsart}
\usepackage{amsfonts}
\usepackage{amssymb}
\usepackage{color}
\usepackage{slashbox}
\usepackage{amssymb,amsmath}
\usepackage{mathrsfs} 
\usepackage{graphicx}

\begin{document}
\begin{frontmatter}
\title{Profitable Bayesian implementation in one-shot mechanism settings}
\author{Haoyang Wu\corauthref{cor}}
\corauth[cor]{Corresponding author.} \ead{18621753457@163.com}
\address{Wan-Dou-Miao Research Lab, Shanghai, China.}

\begin{abstract}
In the mechanism design theory, a designer would like to implement a desired social choice function which specifies her favorite outcome for each possible profile of all agents' types. Traditionally, the designer may be in a dilemma in the sense that even if she is not satisfied with some outcome with low profit, she has to announce it because she must obey the mechanism designed by herself. In this paper, we investigate a case where the designer can induce each agent to adjust his type in a one-shot mechanism. We propose that for a profitable Bayesian implementable social choice function, the designer may escape from the above-mentioned dilemma by spending the optimal adjustment cost and obtain a higher profit. Finally, we construct an example to show that the designer can breakthrough the limit of expected profit which she can obtain at most in the traditional optimal auction model.
\end{abstract}
\end{frontmatter}

\section{Introduction}
In the framework of mechanism design theory \cite{MWG1995, Narahari2009, Serrano2004}, there are one designer and some agents.\footnote{The designer is denoted as ``She'', and the agent is denoted as ``He''.} The designer would like to implement a desired social choice function which specifies her favorite outcome for each possible profile of agents' types. However, agents' types are modelled as their private properties and unknown to the designer. In order to implement the social choice function, the designer constructs a mechanism which specifies each agent's strategy set ($i.e.$, the allowed actions of each agent) and an outcome function ($i.e.$, a rule for how agents' actions get turned into a social choice).

Traditionally, in a standard mechanism agents interact only once ($i.e.$, in one-shot settings), and the designer has no way to adjust agents' types. Hence, the designer may be in a \emph{dilemma} in the sense that even if some profile of agents' strategies leads to an outcome with low profit, she has to announce it because she must obey the mechanism designed by herself. The designer may improve her situation by constructing a multi-period mechanism, or holding a charity auction \cite{Engers2007}. Engers and McManus \cite{Engers2007} proposed that agents' bids in a first-price charity auction are greater than those in a standard (non-charity) auction \cite{Myerson1981} because of the charitable benefit that winners receive from their own payments. Besides the multi-period mechanism and the charity auction, there may exist another way for the designer to escape from the dilemma.

For example, suppose the designer is an auctioneer who wants to sell a good in a hotel, and each agent is a bidder whose initial valuation to the good ($i.e.$, private type) is low. The gorgeousness of the hotel is an open signal to all agents that induces each agent to adjust his valuation to the good before he submits his bid to the designer. Without loss of generality, we assume that each agent's valuation and bid both increase concavely with the hotel rent spent by the designer, and the designer's utility is a linear function of the winner agent's bid. From the viewpoint of the designer, as long as her marginal utility is greater than her marginal rent cost, it is worthwhile for her to continue investing on the rent cost. Obviously, the designer will obtain the maximum profit when her marginal utility is equal to her marginal cost. Thus,  if agents' types ($i.e.$, valuations to the good) are adjustable and influenced by the rent cost of the hotel, the designer may get an outcome better than what would happened without doing so, and \emph{actively} escape from the above-mentioned dilemma.

In this paper, we focus on the one-shot mechanism settings and investigate a case where the designer can induce each agent to adjust his type. In Section 2, we define notions such as adjusted types, optimal adjustment cost, profitable Bayesian implementability and so on. The main result is Proposition 2, $i.e.$, for a profitable Bayesian implementable social choice function, the designer may escape from the above-mentioned dilemma by spending the optimal adjustment cost and obtain a higher profit. In Section 3, we construct an example to show that the designer can breakthrough the limit of expected profit which she can obtain at most in the traditional optimal auction model. Section 4 makes conclusions.

\section{Theoretical analysis}
Following Section 23.B of Mas-Colell, Whinston and Green's textbook \cite{MWG1995}, we consider a one-shot setting with one designer and $I$ agents, indexed by $i=1,\cdots,I$. Let $X$ be a set of possible alternatives.

\emph{Assumption 1:} Each agent $i$ is assumed to observe a private parameter ($i.e.$, \emph{type} $\theta_{i}$) which determines his preference over alternatives in $X$. Let $\Theta_{i}$ be the set of agent $i$'s all possible types. Let
$\Theta=\Theta_{1}\times\cdots\times\Theta_{I}$, $\theta=(\theta_{1}, \cdots, \theta_{I})\in\Theta$.

\textbf{Definition 1}: For any $x\in X$, each \emph{agent $i$'s utility} is denoted as $u_{i}(x, \theta_{i})\in R$, where $\theta_{i}\in\Theta_{i}$, and the \emph{designer's utility} is denoted as $u_{d}(x)\in R$.

According to Ref \cite{MWG1995}, a \emph{social choice function} (SCF) is a function $f:\Theta\rightarrow X$
that, for each possible profile of the agents' types $\theta\in\Theta$, assigns a collective choice $f(\theta)\in X$. A \emph{mechanism} $\Gamma=(S_{1},\cdots,S_{I}$, $g(\cdot))$ is a collection of $I$ strategy sets $S_{1},\cdots,S_{I}$,  and an outcome function $g:S_{1}\times\cdots\times S_{I}\rightarrow X$. A \emph{strategy} of each agent $i$ in $\Gamma$ is a function $s_{i}(\cdot): \Theta_{i} \rightarrow S_{i}$. Let $s(\cdot)=(s_{1}(\cdot), \cdots, s_{I}(\cdot))$.

\emph{Assumption 2:} Assume that in a mechanism, each agent $i$ can play his strategy $s_{i}$ without any cost. Hence agent $i$'s profit with respect to an outcome $x$ is just his utility $u_{i}(x, \theta_{i})$.\footnote{For example, suppose that each agent is a bidder in an auction, then each agent can be considered to submit his bid to the auctioneer without any cost.}  

\emph{Assumption 3:} Assume that in a mechanism, the designer constructs an outcome function and announces a cost $c\geq 0$ which she will spend to perform the outcome function. The cost $c$ is observable to all agents and acts as an open signal. The outcome function is denoted as $g^{c}(\cdot):S_{1}\times\cdots\times S_{I}\rightarrow X$, and the mechanism is denoted as $\Gamma^{c}=(S_{1},\cdots,S_{I}$, $g^{c}(\cdot))$. After learning the cost $c$, each agent $i$ is assumed to adjust his private type from the initial value $\theta_{i}^{0}\in\Theta_{i}$ to a new value $\theta_{i}^{c}\in\Theta_{i}$,\footnote{In Ref \cite{Myerson1981} (Page 60, Line 12), Myerson proposed that ``if there are quality uncertainties, then bidder $i$ might tend to revise his valuation of the object after learning about other bidders' value estimates.'' Similarly, here it is reasonable to assume that each agent $i$ can adjust his private type after observing the cost signal sent by the designer. Section 3 gives an example, where the designer spends some cost to rent a hotel to hold an auction. The gorgeousness of the hotel is just the signal that the designer sends to agents in order to show how precious the sold good is, although the designer may sell a poor good but deliberately rent a luxurious hotel to deceive agents. After observing the signal, each agent adjusts his private type ($i.e.$, valuation to the good).} and then plays strategy $s_{i}(\theta_{i}^{c})$. At last, the designer announces $g^{c}(s_{1}(\theta_{1}^{c}),\cdots, s_{I}(\theta_{I}^{c}))$ as the outcome. The cost $c$ is also denoted as \emph{adjustment cost}. Thus, although the designer does not know each agent's private type exactly, she can induce each agent to adjust his type in a one-shot mechanism.

\textbf{Definition 2:} Given each agent $i$'s initial type $\theta_{i}^{0}\in\Theta_{i}$, for any adjustment cost $c\geq 0$, each agent $i$'s preference over the alternatives in $X$ is determined by his \emph{adjusted type}  $\theta_{i}^{c}\in\Theta_{i}$ by Assumption 3. For each agent $i=1,\cdots, I$, let
\begin{align*}
  &\theta^{0}=(\theta_{1}^{0}, \cdots, \theta_{I}^{0})\in\Theta,\\
  &\theta_{-i}^{0}=(\theta_{1}^{0}, \cdots, \theta_{i-1}^{0}, \theta_{i+1}^{0}, \cdots, \theta_{I}^{0}),\\
  &\theta^{c}=(\theta_{1}^{c}, \cdots, \theta_{I}^{c})\in\Theta,\\
  &\theta_{-i}^{c}=(\theta_{1}^{c}, \cdots, \theta_{i-1}^{c}, \theta_{i+1}^{c}, \cdots, \theta_{I}^{c}).
\end{align*}
 A \emph{type adjustment function} is denoted as $\mu(\theta, c): \Theta\times R^{+}\rightarrow\Theta$, in which $\mu(\theta, 0)=\theta$ for any $\theta\in\Theta$, $i.e.$ zero adjustment cost means no type adjustment. Let $\theta^{c} = \mu(\theta^{0}, c)$. Let $\phi^{0}(\theta^{0}) = (\phi_{1}^{0}(\theta_{1}^{0}), \cdots,\phi_{I}^{0}(\theta_{I}^{0}))$ be the probability density function of initial type profile $\theta^{0}\in\Theta$, and $\phi^{c}(\theta^{c}) = (\phi_{1}^{c}(\theta_{1}^{c}), \cdots,\phi_{I}^{c}(\theta_{I}^{c}))$ be the probability density function of adjusted type profile $\theta^{c}\in\Theta$.
  For each $i=1,\cdots, I$, let
 \begin{align*}
 &\phi_{-i}^{0}(\theta_{-i}^{0}) = (\phi_{1}^{0}(\theta_{1}^{0}), \cdots, \phi_{i-1}^{0}(\theta_{i-1}^{0}), \phi_{i+1}^{0}(\theta_{i+1}^{0}), \cdots,\phi_{I}^{0}(\theta_{I}^{0})),\\
 &\phi_{-i}^{c}(\theta_{-i}^{c}) = (\phi_{1}^{c}(\theta_{1}^{c}), \cdots, \phi_{i-1}^{c}(\theta_{i-1}^{c}), \phi_{i+1}^{c}(\theta_{i+1}^{c}), \cdots,\phi_{I}^{c}(\theta_{I}^{c})).
\end{align*}

\emph{Assumption 4:} For any $\theta\in\Theta$ and adjustment cost $c\geq 0$, the designer is assumed to know the type adjustment function $\mu(\theta, c)$, the initial type distribution $\phi^{0}(\cdot)$ and the adjusted type distribution $\phi^{c}(\cdot)$. \footnote{The initial type profile $\theta^{0}$ and the adjusted type profile $\theta^{c}$ are agents' private information by Assumption 1.}

\textbf{Definition 3}: Given an SCF $f$ and $\phi^{0}(\cdot)$, for any adjustment cost $c\geq 0$, \emph{the designer's expected utility} is denoted as
\begin{equation*}
\bar{u}_{d}(c)=E_{\theta^{c}}u_{d}(f(\theta^{c}))=\int_{\theta^{c}\in\Theta}u_{d}(f(\theta^{c}))\phi^{c}(\theta^{c}) d\theta^{c},
 \end{equation*}
 and \emph{the designer's initial expected utility} is denoted as $\bar{u}_{d}(0)=E_{\theta^{0}}u_{d}(f(\theta^{0}))$.

\textbf{Definition 4}: Given an SCF $f$ and $\phi^{0}(\cdot)$, for any adjustment cost $c\geq 0$, the \emph{designer's expected profit} is denoted as $\bar{p}_{d}(c)=\bar{u}_{d}(c) - c$, and her \emph{initial expected profit} is denoted as $\bar{p}_{d}(0)=\bar{u}_{d}(0)$.

\emph{Assumption 5:} $\bar{u}_{d}(c)$ is assumed to be a concave function with respect to the adjustment cost $c$, $i.e.$,
\begin{equation*}
  \quad \frac{\partial\bar{u}_{d}(c)}{\partial c}>0, \quad \frac{\partial^{2} \bar{u}_{d}(c)}{\partial c^{2}}<0,\quad \text{for any } c\geq 0.\footnote{See the example given in Section 3. Suppose each agent $i$'s adjusted type is a square root function of the designer's cost as specified by Eq (\ref{theta_i}) and the social choice function is specified by Eq (\ref{SCF f}), then the inequalities in Assumption 5 holds.}
\end{equation*}

\textbf{Proposition 1}:   If there exists an adjustment cost $c^{*}\geq 0$  such that
\begin{equation*}
  \frac{\partial \bar{u}_{d}(c)}{\partial c}\Big|_{c=c^{*}}= 1, \quad i.e.\quad\frac{\partial \bar{p}_{d}(c)}{\partial c}\Big|_{c=c^{*}}= 0,
\end{equation*}
then the designer's expected profit $\bar{p}_{d}(c)$ will reach its maximum at $c=c^{*}$. Let $c^{*}$ be denoted as the \emph{optimal adjustment cost}. By Assumption 5, if $c^{*}\geq 0$,  there holds
\begin{equation*}
\frac{\partial \bar{u}_{d}(c)}{\partial c}\Big|_{c=0}\geq 1, \quad i.e.\quad\frac{\partial \bar{p}_{d}(c)}{\partial c}\Big|_{c=0}\geq 0.
 \end{equation*}

According to Ref \cite{MWG1995}, the strategy profile
$s^{*}(\cdot)=(s^{*}_{1}(\cdot),\cdots,s^{*}_{I}(\cdot))$ is a
\emph{Bayesian Nash equilibrium} of mechanism
$\Gamma=(S_{1},\cdots,S_{I},g(\cdot))$ if, for all agent $i$ and all
$\theta_{i}\in\Theta_{i}$, $\hat{s}_{i}\in S_{i}$,
\begin{equation}\label{MWG_BNE}
  E_{\theta_{-i}}[u_{i}(g(s^{*}_{i}(\theta_{i}),s^{*}_{-i}(\theta_{-i})),\theta_{i})|\theta_{i}]
  \geq
  E_{\theta_{-i}}[u_{i}(g(\hat{s}_{i},s^{*}_{-i}(\theta_{-i})),\theta_{i})|\theta_{i}].
\end{equation}
The mechanism
$\Gamma=(S_{1},\cdots,S_{I},g(\cdot))$ \emph{implements the social choice
function} $f(\cdot)$ \emph{in Bayesian Nash equilibrium} if there is a
Bayesian Nash equilibrium of $\Gamma$,
$s^{*}(\cdot)=(s^{*}_{1}(\cdot),\cdots,s^{*}_{I}(\cdot))$, such that
$g(s^{*}(\theta))=f(\theta)$ for all $\theta\in\Theta$.

\textbf{Definition 5}: Given an SCF $f$ and $\phi^{0}(\cdot)$,
 $f$ is \emph{profitable Bayesian implementable} if the following conditions are satisfied:\\
 1)  The optimal adjustment cost $c^{*}>0$. \footnote{Hence, the distribution of agents' private types will be adjusted from $\phi^{0}(\cdot)$ to $\phi^{c^{*}}(\cdot)$ after each agent observes $c^{*}$.}\\
 2)  There exist a mechanism $\Gamma^{c^{*}}=(S_{1},\cdots,S_{I},g^{c^{*}}(\cdot))$ that implements $f$ in Bayesian Nash equilibrium. That is, there exists
 a strategy profile $s^{*}(\cdot)=(s^{*}_{1}(\cdot),\cdots,s^{*}_{I}(\cdot))$ such that:\\
(i) For all agent $i$ and all $\theta_{i}^{c^{*}}\in\Theta_{i}$,
\begin{equation}\label{profitable_BNE}
  E_{\theta_{-i}^{c^{*}}}[u_{i}(g^{c^{*}}(s^{*}_{i}(\theta_{i}^{c^{*}}),s^{*}_{-i}(\theta_{-i}^{c^{*}})), \theta_{i}^{c^{*}})|\theta_{i}^{c^{*}}]
  \geq
  E_{\theta_{-i}^{c^{*}}}[u_{i}(g^{c^{*}}(\hat{s}_{i},s^{*}_{-i}(\theta_{-i}^{c^{*}})), \theta_{i}^{c^{*}})|\theta_{i}^{c^{*}}]
\end{equation}
for all $\hat{s}_{i}\in S_{i}$. \footnote{\label{Footnote7}Note that in formula (\ref{profitable_BNE}), the probability density function of type profile $\theta_{-i}^{c^{*}}=(\theta_{1}^{c^{*}}, \cdots, \theta_{i-1}^{c^{*}}, \theta_{i+1}^{c^{*}}, \cdots, \theta_{I}^{c^{*}})$ is $\phi_{-i}^{c^{*}}(\cdot)$. As a comparison, in the conventional notion of Bayesian Nash equilibrium, there is no type adjustment, and the probability density function of type profile $\theta_{-i}=(\theta_{1}, \cdots, \theta_{i-1}, \theta_{i+1}, \cdots, \theta_{I})$ in formula (\ref{MWG_BNE}) is $\phi_{-i}^{0}(\cdot)$.}
\\(ii) $g^{c^{*}}(s^{*}(\theta))=f(\theta)$ for all $\theta\in\Theta$.

\textbf{Proposition 2}: Given an SCF $f$ and $\phi^{0}(\cdot)$, if $f$ is profitable Bayesian implementable, then by spending the optimal adjustment cost, the designer can obtain an expected profit larger than her initial expected profit.

\textbf{Proof}: Given that $f$ is profitable Bayesian implementable, then the optimal adjustment cost $c^{*}>0$.
By Proposition 1, $\bar{p}_{d}(c^{*})$ is the maximum expected profit. Therefore, $\bar{p}_{d}(c^{*})>\bar{p}_{d}(0)$. $\Box$

According to Ref \cite{MWG1995},  a social choice function $f(\cdot)$ is \emph{truthfully implementable in Bayesian Nash equilibrium} (or \emph{Bayesian incentive compatible}) if $s^{*}_{i}(\theta_{i})=\theta_{i}$ for all $\theta_{i}\in\Theta_{i}$ ($i=1,\cdots,I$) is a Bayesian Nash equilibrium of the direct mechanism $\Gamma=(S_{1},\cdots,S_{I},g(\cdot))$, in which $S_{i}=\Theta_{i}$, $g=f$. That is, if for all $i=1,\cdots,I$ and all $\theta_{i}\in\Theta_{i}$, $\hat{\theta}_{i}\in \Theta_{i}$,
\begin{equation}\label{MWG_BIC}
  E_{\theta_{-i}}[u_{i}(f(\theta_{i},\theta_{-i}),\theta_{i})|\theta_{i}]
 \geq
  E_{\theta_{-i}}[u_{i}(f(\hat{\theta}_{i},\theta_{-i}),\theta_{i})|\theta_{i}].
\end{equation}

Proposition 23.D.1 \cite{MWG1995}: (\emph{The Revelation Principle for Bayesian Nash Equilibrium}) Suppose that there exists a mechanism $\Gamma=(S_{1},\cdots,S_{I},g(\cdot))$ that implements the social choice function $f(\cdot)$ in Bayesian Nash equilibrium. Then $f(\cdot)$
is truthfully implementable in Bayesian Nash equilibrium.

\textbf{Proposition 3}: Given an SCF $f$ and $\phi^{0}(\cdot)$, if $f$ is profitable Bayesian implementable, then it cannot be inferred that $f$ is truthfully implementable in Bayesian Nash equilibrium. That is, the revelation principle does not hold in this  case.

\textbf{Proof}: Given that $f$ is profitable Bayesian implementable, then the optimal adjustment cost $c^{*}>0$, and there exist a mechanism $\Gamma^{c^{*}}=(S_{1},\cdots,S_{I},g^{c^{*}}(\cdot))$ that implements $f$ in Bayesian Nash equilibrium. According to Footnote \ref{Footnote7}, formula (\ref{profitable_BNE}) is related to the type distribution $\phi^{c^{*}}(\cdot)$. \\
As a comparison, it can be seen from formula (\ref{MWG_BIC}) that in the notion of Bayesian incentive compatibility,  there is no type adjustment in the direct mechanism. Thus, formula (\ref{MWG_BIC}) is related to the type distribution $\phi^{0}(\cdot)$.\\
Since $\theta^{c^{*}} = \mu(\theta^{0}, c^{*})$ and $c^{*}>0$, thus $\phi^{0}(\cdot)$ is not equal to $\phi^{c^{*}}(\cdot)$. Obviously, formula (\ref{MWG_BIC}) cannot be inferred from formula (\ref{profitable_BNE}). Therefore, given that $f$ is profitable Bayesian implementable, it cannot be inferred that $f$ is truthfully implementable in Bayesian Nash equilibrium. Consequently, the revelation principle for Bayesian Nash Equilibrium does not hold in this case. $\Box$

\textbf{Proposition 4}: If the designer's expected utility $\bar{u}_{d}(c)$ and expected profit $\bar{p}_{d}(c)$ satisfy the following condition,
\begin{equation}
\frac{\partial\bar{u}_{d}(c)}{\partial c}\Big|_{c=0}< 1,  \quad i.e.\quad\frac{\partial \bar{p}_{d}(c)}{\partial c}\Big|_{c=0}< 0,
\end{equation}
then the designer will obtain the maximum expected profit at $c=0$. Put differently, in this case the designer cannot obtain any more profit by spending any cost to induce each agent to adjust his type.

\section{Example}
\subsection{Model Settings}
Following the auction model in MWG's book (Page 863, \cite{MWG1995}), suppose that there are one designer and two agents. Let the designer be an auctioneer who wants to sell a good, and each agent be a bidder whose valuation to the good is $\theta_{i}\geq 0$, $i.e.$, $\Theta_{i}=R^{+}$. We consider a first-price-sealed-bid auction setting: Each agent $i$ is allowed to submit a sealed bid $b_{i}\geq 0$. The bids are then opened, and the agent with the higher bid gets the good, and must pay money equal to his bid to the auctioneer.

Suppose that:\\
1) Each agent $i$'s initial valuation ($i.e.$, his initial type) $\theta_{i}^{0}$ is drawn independently from the uniform distribution on $[0,1]$. The distribution is known by the designer but the exact value of each $\theta_{i}^{0}$ is agent $i$'s private information.

2) The designer holds the auction in a hotel, and the cost for renting the hotel is $c\geq 0$, which is observable to two agents.

3) The gorgeousness of hotel is characterized by the rent cost $c$. Each agent $i$ adjusts his private valuation to the good after observing the gorgeousness of the hotel. The larger the rent cost is, the greater each bidder's private valuation to the good will be.

4) Let $\beta>0$ be a coefficient, each agent $i$'s valuation to the good ($i.e.$, his adjusted type $\theta_{i}^{c}$) is a square root function of the rent cost $c$,
\begin{equation}\label{theta_i}
\theta_{i}^{c}= (1+\beta\sqrt{c})\theta_{i}^{0}.
\end{equation}
Thus,
\begin{equation*}
\frac{\partial\theta_{i}^{c}}{\partial c}=\frac{\beta\theta_{i}^{0}}{2\sqrt{c}}, \quad\text{ } \frac{\partial^{2}\theta_{i}^{c}}{\partial c^{2}}=-\frac{\beta\theta_{i}^{0}}{4}c^{-3/2}.
\end{equation*}
 That is, for any $c\geq0$, the following formulas hold:
\begin{equation*}
    \frac{\partial\theta_{i}^{c}}{\partial c}\Big|_{c=0}=+\infty, \text { } \frac{\partial\theta_{i}^{c}}{\partial c}>0, \text { }\frac{\partial^{2}\theta_{i}^{c}}{\partial c^{2}}<0.
\end{equation*}
Let $\theta=(\theta_{1},\theta_{2})$, consider the social choice function
\begin{equation}\label{SCF f}
f(\theta)=(y_{1}(\theta), y_{2}(\theta), y_{d}(\theta), t_{1}(\theta), t_{2}(\theta), t_{d}(\theta)),
\end{equation}
in which
\begin{align*}
&y_{1}(\theta)=1, \quad \text{if }\theta_{1}\geq\theta_{2};\quad =0 \text{ if }\theta_{1}<\theta_{2}\\
&y_{2}(\theta)=1, \quad \text{if }\theta_{1}<\theta_{2};\quad =0 \text{ if }\theta_{1}\geq\theta_{2}\\
&y_{d}(\theta)=0, \quad\text{for all }\theta\in\Theta\\
&t_{1}(\theta)= - \theta_{1} y_{1}(\theta)/2\\
&t_{2}(\theta)= - \theta_{2}y_{2}(\theta)/2\\
&t_{d}(\theta)= [\theta_{1}y_{1}(\theta)+\theta_{2} y_{2}(\theta)]/2.
\end{align*}
The subscript $d$ stands for the designer, and the subscript $1,2$ stands for the agent 1 and agent 2 respectively. $y_{i}=1$ means that agent $i=1, 2$ gets the good. $t_{i}$ denotes agent $i$'s payment to the designer. $t_{d}$ denotes the sum of two agents' payment to the designer. Obviously, for any $c\geq 0$, $y_{1}(\theta^{c})=y_{1}(\theta^{0})$, $y_{2}(\theta^{c})=y_{2}(\theta^{0})$.

\subsection{f is Bayesian Nash implementable}
Now we investigate whether the social choice function $f(\theta)$ is Bayesian Nash implementable. We will look for an equilibrium in which each agent $i$'s strategy $b_{i}(\cdot)$ takes the form $b_{i}(\theta_{i}^{c})=\alpha_{i}\theta_{i}^{c}=\alpha_{i}(1+\beta\sqrt{c})\theta_{i}^{0}$ for $\alpha_{i}\in[0,1]$.

Suppose that agent 2's strategy has this form, and consider agent 1's problem. For each possible $\theta_{1}^{c}$, agent 1 wants to solve the following problem:
\begin{equation*}
  \max\limits_{b_{1}\geq 0}(\theta_{1}^{c} - b_{1}) \text{Prob}(b_{2}(\theta_{2}^{c})\leq b_{1}).
\end{equation*}
Because agent 2's highest possible bid is $\alpha_{2}(1+\beta\sqrt{c})$  when $\theta_{2}^{0}=1$, it is evident that agent 1's bid $b_{1}$ should never more than $\alpha_{2}(1+\beta\sqrt{c})$. Note that $\theta_{2}^{0}$ is uniformly distributed on $[0,1]$, and $b_{2}(\theta_{2}^{c})=\alpha_{2}(1+\beta\sqrt{c})\theta_{2}^{0}\leq b_{1}$ if and only if $\theta_{2}^{0}\leq b_{1}/[\alpha_{2}(1+\beta\sqrt{c})]$. Thus,
\begin{equation*}
\text{Prob}(b_{2}(\theta_{2}^{c})\leq b_{1}) = \frac{b_{1}}{\alpha_{2}(1+\beta\sqrt{c})}.
\end{equation*}
We can write agent 1's problem as:
\begin{equation*}
  \max\limits_{0\leq b_{1}\leq \alpha_{2}(1+\beta\sqrt{c})}\frac{(\theta_{1}^{c} - b_{1}) b_{1}}{\alpha_{2}(1+\beta\sqrt{c})}
\end{equation*}
The solution to this problem is
\begin{equation*}
  b_{1}^{*}(\theta_{1}^{c})
  =\begin{cases}
     \theta_{1}^{c}/2,& \text{if } \theta_{1}^{0}/2\leq\alpha_{2}\\
      \alpha_{2}(1+\beta\sqrt{c}),& \text{if } \theta_{1}^{0}/2>\alpha_{2}
   \end{cases}.
\end{equation*}
Similarly,
\begin{equation*}
  b_{2}^{*}(\theta_{2}^{c})
  =\begin{cases}
     \theta_{2}^{c}/2,& \text{if } \theta_{2}^{0}/2\leq\alpha_{1}\\
      \alpha_{1}(1+\beta\sqrt{c}),& \text{if } \theta_{2}^{0}/2>\alpha_{1}
   \end{cases}.
\end{equation*}
Let $\alpha_{1}=\alpha_{2}=1/2$, we see that the strategies $ b_{i}^{*}(\theta_{i}^{c}) = \theta_{i}^{c}/2$ for $i=1, 2$ constitute a Bayesian Nash equilibrium for this mechanism. Thus, there is a Bayesian Nash equilibrium of this first-price-sealed-bid auction mechanism that indirectly yields the outcomes specified by the social choice function $f(\theta)$, and hence $f(\theta)$ is Bayesian Nash implementable.

\subsection{The designer's expected profit}
Let us consider the designer's expected profit:
\begin{equation*}
 \bar{p}_{d}(c)=(1+\beta\sqrt{c})E[\theta_{1}^{0}y_{1}(\theta^{0})+\theta_{2}^{0}y_{2}(\theta^{0})]/2 - c.
\end{equation*}
The designer's problem is to choose an optimal adjustment cost  $c\geq0$ to maximize her expected profit, $i.e.$,
\begin{equation*}
  \max\limits_{c\geq0}(1+\beta\sqrt{c})E[\theta_{1}^{0}y_{1}(\theta^{0})+\theta_{2}^{0}y_{2}(\theta^{0})]/2 - c
\end{equation*}
By appendix, the designer's initial expected profit is $\bar{p}_{d}(0)=E[\theta_{1}^{0}y_{1}(\theta^{0})+\theta_{2}^{0}y_{2}(\theta^{0})]/2=1/3$. Thus, the designer's problem is reformulated as:
\begin{equation*}
  \max\limits_{c\geq0}(1+\beta\sqrt{c})/3 - c
\end{equation*}
It can be easily derived that the optimal adjustment cost $c^{*}=\beta^{2}/36$.  Hence, by Definition 5 $f(\theta)$ is profitable Bayesian implementable. The maximum expected profit of the designer is:
\begin{equation*}
  \bar{p}_{d}(c^{*})=(1+\beta\sqrt{c^{*}})/3 - c^{*} = \frac{1}{3}(1+\frac{\beta^{2}}{12}).
\end{equation*}
Obviously, when $\beta>\sqrt{3}$, there exists $\bar{p}_{d}(c^{*})>5/12$. Note that the designer's maximum expected profit in the traditional optimal auction with two bidders is $5/12$ (see Page 23, the ninth line from the bottom, Ref \cite{Krishna2010}). Therefore, if $\beta>\sqrt{3}$, then by choosing the optimal adjustment cost $c^{*}=\beta^{2}/36$, \emph{the designer can breakthough the limit of expected profit which she can obtain at most in the traditional optimal auction model}.

\subsection{Each agent's ex ante expected profit}
Now we consider each agent's \emph{ex ante} expected profit when agents' types are adjustable and the designer chooses the optimal adjustment cost $c^{*}$. By appendix, the winner agent's expected profit is denoted as follows:
\begin{align*}
E[\theta_{winner}^{c^{*}} - b_{winner}^{*}(\theta_{winner}^{c^{*}})] &= E[\theta_{winner}^{c^{*}}/2] =(1+\beta\sqrt{c^{*}})E[\theta_{winner}^{0}]/2\\
&=(1+\beta\sqrt{c^{*}})E[\theta_{1}^{0}y_{1}(\theta^{0})+\theta_{2}^{0}y_{2}(\theta^{0})]/2\\
&=\frac{1}{3}(1+\frac{\beta^{2}}{6}).
\end{align*}
And the loser agent's expected profit is zero. Because the two agents are symmetric, each of them has the same probability $1/2$ to be the winner agent. Therefore, each agent's \emph{ex ante} expected profit is $1/6+\beta^{2}/36$.

As a comparison, we consider the \emph{ex ante} expected profit of each agent in the traditional optimal auction model.
By Ref \cite{Krishna2010} (Page 22), the \emph{ex ante} expected payment of a bidder is
\begin{equation*}
  r(1-F(r))G(r)+\int_{r}^{\omega}y(1-F(y))g(y)dy,
\end{equation*}
where  $r>0$ is the reserve price, $[r, \omega]$ is the interval of each agent $i$'s valuation $X_{i}$, which is independently and identically distributed according to the increasing distribution function $F$. Fix a bidder, $G$ denotes the distribution function of the highest valuation among the rest remaining bidders.
For the case of two agents with valuation range $[r, 1]$ and uniform distribution,
\begin{equation*}
  F(r)=r,\text{ }G(r)=r, \text{ }\omega=1,\text{ }F(y)=y, \text{ }g(y)=1, \text{for any } y\in[r, 1].
\end{equation*}
By Ref \cite{Krishna2010} (Page 23), given that each of two agent's valuation to the good is uniformly distributed on interval $[0, 1]$, the optimal reserve price $r^{*}=1/2$. Therefore, the \emph{ex ante} expected payment of each agent in the traditional optimal auction is
\begin{align*}
r^{*}&(1-r^{*})r^{*}+\int_{r^{*}}^{1}y(1-y)dy\\
&=\frac{1}{8}+\int_{\frac{1}{2}}^{1}y(1-y)dy=\frac{5}{24}.
\end{align*}
Since the optimal reserve price is $1/2$, each agent's valuation to the good is uniformly distributed on interval $[1/2, 1]$. Hence, each agent's expected valuation is the middle point of interval $[1/2, 1]$, $i.e.$, $3/4$. Consequently, the \emph{ex ante} expected profit of each agent in the traditional optimal auction is his expected valuation $3/4$ minus his \emph{ex ante} expected payment $5/24$, $i.e.$,
\begin{equation}
 \frac{3}{4}-\frac{5}{24}=\frac{13}{24}.
\end{equation}
Recall that when agents' types are adjustable and the designer chooses the optimal adjustment cost $c^{*}=\beta^{2}/36$, each agent's \emph{ex ante} expected profit is $1/6+\beta^{2}/36$. It can be seen that if $\beta>\sqrt{27/2}$, then $1/6+\beta^{2}/36 > 13/24$.


\section{Conclusions}
In the standard mechanism design theory, the mechanism works in a one-shot manner. Each agent's type is considered as private and endogenous value, which means that the designer has no way to know and adjust each agent's type. Thus, although the designer constructs a mechanism in order to implement her favorite social choice function, she may behave like \emph{a passive observer in a dilemma} after receiving a profile of agents' strategies: $i.e.$, she must obey the mechanism and announce the outcome specified by the outcome function, even if she is not satisfied with the outcome.

In this paper, we investigate a case where the designer can induce each agent to adjust his type just in a one-shot mechanism. The novelties of this paper are as follows:\\
1) As shown in Proposition 2, for a profitable Bayesian implementable social choice function, the designer may behave like \emph{an active modulator} and escape from the above-mentioned dilemma by spending the optimal adjustment cost and obtain a higher profit. \\
2) As shown in Proposition 3, it cannot be inferred that a profitable Bayesian implementable social choice function is truthfully implementable in Bayesian Nash equilibrium. That is, the revelation principle does not hold in this case.\\
3) As shown in Section 3, the designer can breakthrough the limit of expected profit which she can obtain at most in the traditional optimal auction model:\\
$\bullet$ If $\beta>\sqrt{3}$, then by choosing the optimal adjustment cost $c^{*}=\beta^{2}/36$, the designer can obtain an expected profit larger than the maximum expected profit $5/12$ yielded by the traditional optimal auction. \\
$\bullet$ If $\beta>\sqrt{27/2}$, each agent's \emph{ex ante} expected profit is also larger than the corresponding value $13/24$ in the traditional optimal auction.

\section*{Appendix}
As specified in Section 3, $\theta_{1}^{0}$ and $\theta_{2}^{0}$ are drawn independently from the uniform distribution on $[0,1]$. Let $Z$ be a random variable $Z=\theta_{1}^{0}y_{1}(\theta^{0})+\theta_{2}^{0}y_{2}(\theta^{0})$.
\begin{equation*}
  f_{\theta_{1}^{0}}(z)
  =\begin{cases}
      0,& z<0\\
      1,& z\in[0,1]\\
      0,& z>1
   \end{cases}.
\end{equation*}
\begin{equation*}
  F_{\theta_{1}^{0}}(z)=Prob\{\theta_{1}^{0}\leq z\}
  =\begin{cases}
      0,& z<0\\
      z,& z\in[0,1]\\
      1,& z>1
   \end{cases}.
\end{equation*}
\begin{equation*}
  F_{Z}(z) =  [F_{\theta_{1}^{0}}(z)]^2
  =\begin{cases}
      0,& z<0\\
      z^2,& z\in[0,1]\\
      1,& z>1
   \end{cases}.
\end{equation*}
Therefore,
\begin{equation*}
  f_{Z}(z)
  =\begin{cases}
      0,& z<0\\
      2z,& z\in[0,1]\\
      0,& z>1
   \end{cases}.
\end{equation*}
As a result,
\begin{equation*}
  E(Z)=\int_{0}^{1}z\cdot 2zdz=\int_{0}^{1}2z^2dz=2/3.
\end{equation*}
Therefore, $E[\theta_{1}^{0}y_{1}(\theta^{0})+\theta_{2}^{0}y_{2}(\theta^{0})]/2 = 1/3$. According to Eq (\ref{SCF f}), the designer's initial expected profit and utility are $\bar{p}_{d}(0)=\bar{u}_{d}(0)=E[\theta_{1}^{0}y_{1}(\theta^{0})+\theta_{2}^{0}y_{2}(\theta^{0})]/2=1/3$.

\section*{Acknowledgments}
The author is grateful to Fang Chen, Hanyue Wu, Hanxing Wu and Hanchen Wu for their great support.

\end{document}